\documentstyle[aps,prb,graphicx,multicol]{revtex}

\def\bs#1{\bbox{#1}}
\def\ve#1{\bs{{\mathbf{#1}}}}

\newcommand{\be}{\begin{eqnarray}}
\newcommand{\ee}{\end{eqnarray}}
\newcommand{\nn}{\nonumber}

\begin{document}

\title{Generalization of the coupled dipole method to periodic structures}

\author{Patrick C. Chaumet} \address{Institut Fresnel (Unit\'e Mixte
de Recherche 6133), Facult\'e des Sciences et Techniques de St
J\'er\^ome, Av. Escadrille Normandie-Niemen, F-13397 Marseille cedex
20, France}

\author{Adel Rahmani}

\address{Laboratoire d'Electronique, Opto\'electronique et
Microsyst\`emes, UMR CNRS ECL 5512,\\ 36 avenue Guy de Collongue, BP 163,
F-69131 Ecully, France}

\author{Garnett W. Bryant}

\address{Atomic Physics Division, National Institute of Standards and
Technology, Gaithersburg, Maryland 20899-8423}

\date{\today}

\maketitle

\begin{abstract}
We present a generalization of the coupled dipole method to the
scattering of light by arbitrary periodic structures. This new
formulation of the coupled dipole method relies on the same
direct-space discretization scheme that is widely used to study the
scattering of light by finite objects. Therefore, all the knowledge
acquired previously for finite systems can be transposed to the study
of periodic structures.
\end{abstract}

\begin{multicols}{2}

\section{Introduction}
In its original form, the coupled dipole method (CDM) was developed
for the study, in free-space, of the scattering of light by an object with
finite dimensions.\cite{purcell:73,draine:88} The method was
subsequently extended to deal with objects near a
substrate~\cite{schmehl:97,chaumet1} or inside a multilayer
system.\cite{rahmani1} The principle of the method is always the same:
the object is represented by a cubic array of $N$ polarizable
subunits, each with a size small enough compared to the spatial
variations of the electromagnetic field for the dipole approximation
to apply.  If the CDM could be extended to deal with local scatterers
near periodic structures, the CDM could then also be used, for
example, to study light scattering by objects near surface gratings or
by defects or cavities in photonic crystals. The first step toward
such an extension is to develop a form of the CDM capable of
describing periodic structures efficiently.  In this paper, we present
a generalization of the CDM to arbitrary periodic structures.

\section{Self-consistent field for a periodic structure}
We consider a plane substrate occupying the region $z \leq 0$.  For a
single object on the substrate, the self-consistent field at the
$i^{\rm th}$ subunit at location $\ve{r}_i$ is given by
\be\label{dipi1}
\label{dipi} \ve{E}(\ve{r}_i,\omega) & = &
\ve{E}_0(\ve{r}_i,\omega) + \sum_{j=1}^{N} [
\ve{S}(\ve{r}_i,\ve{r}_j,\omega)\nonumber\\ & + &
\ve{F}(\ve{r}_i,\ve{r}_j,\omega)] \alpha_j(\omega)
\ve{E}(\ve{r}_j,\omega). \ee
where $\ve{E}_0(\ve{r}_i,\omega)$ is the (initial) field at $\ve{r}_i$
in the absence of the scattering object. Note that none of the
subunits lies in the plane $z=0$. The tensors $\ve{F}$ and $\ve{S}$
are the field susceptibilities (linear responses) associated with the
free space~\cite{rahmani2} and the substrate.~\cite{agarwal}
$\alpha_i(\omega)$ is the dynamic polarizability of the $i^{th}$
subunit and includes radiation reaction.~\cite{draine:88,chaumet2} The
self-consistent field $\ve{E}(\ve{r}_i,\omega)$ is found by solving
the symmetric linear system formed by writing Eq.~(\ref{dipi}) for
$i=1,N$.  The total field at position $\ve{r}$ is computed as
	
\be\label{dipi2} \ve{E}(\ve{r},\omega)=\ve{E}_0(\ve{r},\omega) &
+ & \sum_{j=1}^{N}[ \ve{S}(\ve{r},\ve{r}_j,\omega)\nonumber\\ & + &
\ve{F}(\ve{r},\ve{r}_j,\omega)] \alpha  _j(\omega)
\ve{E}(\ve{r}_j,\omega).\ee

This conventional form of the CDM is well adapted to deal with
localized objects. If, instead of a single object, one wants to study
a periodic structure created by the repetition of the object over a
lattice located above the substrate, Eq.~(\ref{dipi}) becomes

\be\label{dipi3}&&\ve{E}(\ve{r}_i,\omega)=\ve{E}_0(\ve{r}_i,\omega)
\nonumber\\ & + & \sum_{j=1}^{N}
\sum_{m,n=-\infty}^{\infty}
[\ve{S}(\ve{r}_i,\ve{\bar{r}}_j+m\ve{u}+n\ve{v},\omega)\nonumber\\&+&
\ve{F}(\ve{r}_i,\ve{\bar{r}}_j+m\ve{u}+n\ve{v},\omega)]
\alpha_j(\omega)\ve{E}(\ve{\bar{r}}_j+m\ve{u}+n\ve{v},\omega).\ee 
\begin{figure}[H]
\begin{center}
\resizebox{80mm}{!}{\input{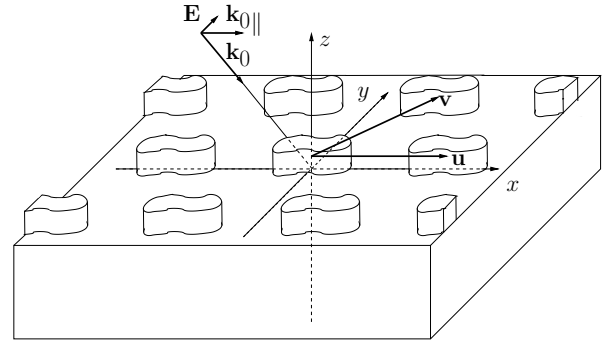}}
\end{center}
\caption{Example of a periodic structure created by the repetition of
an object over a lattice parallel to a substrate.}
\label{figjonction}
\end{figure}
The vectors $\ve{u}$ and $\ve{v}$ are the basis vectors of the lattice
(Fig. 1).  The index $i$ runs over all the subunits of the
structure. $\ve{r}_i$ is the position of subunit $i$. The sum over $j$
is restricted to the $N$ subunits of a single object with position
$\ve{\bar{r}}_j$ inside the object.  The number of subunits is now
infinite, and therefore so is the size of the linear system to be
solved. One solution would be to truncate the infinite sum and solve
the system for a large but finite number of objects, but this is
impractical because the sums over the lattice converge very slowly.
This problem can be circumvented by using a plane-wave decomposition
of the incident field.  In the case of plane-wave (propagating or
evanescent) illumination, the field above the surface can be written
as (we note by $\ve{k}_{0\parallel}$ the projection of vector
$\ve{k}_0$ on a plane parallel to the surface)

\be\label{phase} \ve{E}_0(\ve{\bar{r}}_i+m\ve{u}+n\ve{v},\omega)& = &
\ve{E}_0(\ve{\bar{r}}_i,\omega)
\exp[i\ve{k}_{0\parallel}.(m\ve{u}+n\ve{v})]\nonumber\\ & & \ee
where $\ve{k}_0$ is the wave vector in free space.
Because of
the periodicity of the system and the translational invariance of the
field susceptibilities, the self consistent field satisfies the same
relation as the incident field (Eq.~(\ref{phase})), and at any subunit
Eq.~(\ref{dipi3}) can be written as
\be\label{dipi4}&&\ve{E}(\ve{r}_i,\omega)=\ve{E}_0(\ve{r}_i,\omega)
\nonumber\\ & + & \sum_{j=1}^{N}
\Bigg(\sum_{m,n=-\infty}^{\infty}
[\ve{S}(\ve{r}_i,\ve{\bar{r}}_j+m\ve{u}+n\ve{v},\omega)\nonumber\\&+&
\ve{F}(\ve{r}_i,\ve{\bar{r}}_j+m\ve{u}+n\ve{v},\omega)] \exp[i
\ve{k}_{0\parallel}.(m\ve{u}+n\ve{v})]\Bigg) \nonumber\\ &
&\alpha_j(\omega)\ve{E}(\ve{\bar{r}}_j,\omega).
\ee 
The self-consistent field on the right-hand side of Eq.(\ref{dipi4})
is independent of ($m$,$n$) and can be taken out of the infinite
sum. Hence the sum over subunits in Eq.(\ref{dipi4}) only involves
$j=1,N$, that is the number of subunits in a unit cell, which we choose to be
the cell for which $m=n=0$.  Moreover, because of the translational
symmetry of the self-consistent field, we only need to find $\ve{E}$
in one cell.  Once the self-consistent field is found in the central
cell, the field in any other cell is obtained by multiplying by the
appropriate phase factor.  Thus we only have to solve a linear system
of the same size as the one describing a single object.  The major
issue in solving Eq.~(\ref{dipi4}) is to compute efficiently the
infinite, slowly convergent sums without performing a truncation of
the sums. This is possible owing to the translational invariance of
the field-susceptibilities in a plane parallel to the surface. The
dependence on $(\ve{\bar{r}}_i,\ve{\bar{r}}_j,\omega)$ can be written
as $(\ve{\rho}_{ij},z_i,z_j,\omega)$ with
$\ve{\rho}_{ij}=(\ve{\bar{r}}_i-\ve{\bar{r}}_j)_{\parallel}$.  Hence,
the infinite sums of Eq.(\ref{dipi4}) become:


\be\label{somme} K & = & \sum_{m,n=-\infty}^{\infty}
[\ve{S}(\ve{\bar{r}}_i,\ve{\bar{r}}_j+m\ve{u}+n\ve{v},\omega)
\nonumber\\\nonumber &+&
\ve{F}(\ve{\bar{r}}_i,\ve{\bar{r}}_j+m\ve{u}+n\ve{v},\omega)]\exp[i
\ve{k}_{0\parallel}.(m\ve{u}+n\ve{v})]\\\nonumber &=&\int
d\ve{r}_{\parallel}\sum_{m,n=-\infty}^{\infty}
\delta(\ve{r}_{\parallel}-m\ve{u}-n\ve{v})\exp(i
\ve{k}_{0\parallel}.\ve{r}_{\parallel}) \\ &\times
&[\ve{S}(\ve{\rho}_{ij}-\ve{r}_{\parallel},z_i,z_j,\omega)
+\ve{F}(\ve{\rho}_{ij}-\ve{r}_{\parallel},z_i,z_j,\omega)] \ee

We define the two-dimensional Fourier transform as~:
${\cal{F}}[b(\ve{r}_{\parallel})]=\int d\ve{r}_{\parallel}
b(\ve{r}_{\parallel}) \exp(-i\ve{r}_{\parallel}.\ve{h}_{\parallel})$,
and its inverse as ${\cal{F}}^{-1}[B(\ve{h}_{\parallel})]=1/(2\pi)^2\times
\int d\ve{h}_{\parallel} B(\ve{h}_{\parallel})
\exp(i\ve{r}_{\parallel}.\ve{h}_{\parallel})$.  Using the
Parseval-Plancherel theorem Eq.(\ref{somme}) becomes

\be
\label{dipi7}K & = & \frac{1}{(2\pi)^2}\int
 d\ve{h}_{\parallel} M \sum_{m,n=-\infty}^{\infty}
\delta(\ve{h}_{\parallel}-m\ve{u}'-n\ve{v}'+\ve{k}_{0\parallel})
\nonumber\\ & \times &
{\cal{F}}[\ve{S}(\ve{\rho}_{ij}-\ve{r}_{\parallel},z_i,z_j,\omega)
+\ve{F}(\ve{\rho}_{ij}-\ve{r}_{\parallel},z_i,z_j,\omega)] \ee
where $\ve{u}'=2\pi(v_y\ve{\hat{x}}-v_x\ve{\hat{y}})/(u_xv_y-v_xu_y)$
and $\ve{v}'=2\pi(-u_y\ve{\hat{x}}+u_x\ve{\hat{y}})/(u_xv_y-v_xu_y)$
are the basis vectors of the reciprocal lattice, and
$M=(2\pi)^2/(u_xv_y-v_xu_y)$. $\ve{\hat{x}}$ and $\ve{\hat{y}}$ are
the basis vectors of the coordinate system. Using the angular spectrum
representations $\ve{W}$ and $\ve{G}$ of tensors $\ve{S}$ and $\ve{F}$
Eq.(\ref{dipi7}) becomes\cite{rahmani2,agarwal}
\be\label{dipi8} K & = & \frac{i}{2 \pi}M
\sum_{m,n=-\infty}^{\infty}
\exp[i(m\ve{u}'+n\ve{v}'+\ve{k}_{0\parallel}).\ve{\rho}_{ij}]\times
\nonumber \\\nonumber &&
\{\ve{W}(m\ve{u}'+n\ve{v}'+\ve{k}_{0\parallel},\ve{k}_0)\exp[iw_0(z_i+z_j)]
\\& + & \ve{G}(m\ve{u}'+n\ve{v}'+\ve{k}_{0\parallel},\ve{k}_0)
\exp[iw_0|z_i-z_j|]\}, \ee
with
\be \ve{G}(\ve{k}_{\parallel},\ve{k}_0)  = \left(\begin{array}{ccc}
\frac{k_0^2-k_x^2}{w_0} & -\frac{k_xk_y}{w_0} & -\gamma k_x \\
-\frac{k_xk_y}{w_0} & \frac{k_0^2-k_y^2}{w_0} & -\gamma k_y \\ -\gamma
k_x & -\gamma k_y & \frac{k^2_{\parallel}}{w_0}
\end{array}\right),
\ee
and
\be
& & \ve{W}(\ve{k}_{\parallel},\ve{k}_0)  =  \nn\\
& & 
\left(\begin{array}{ccc}
\frac{k_x^2w_0\Delta_p}{k^2_{\parallel}}-
\frac{k_y^2k_0^2\Delta_s}{k^2_{\parallel} w_0} &
\frac{k_xk_y}{w_0 k^2_{\parallel}}\left(w_0^2\Delta_p+
k_0^2\Delta_s\right) & k_x\Delta_p \\
\frac{k_xk_y}{w_0 k^2_{\parallel}}\left(w_0^2\Delta_p+
k_0^2\Delta_s\right) &
\frac{k_y^2w_0\Delta_p}{k^2_{\parallel}}-
\frac{k_x^2k_0^2\Delta_s}{k^2_{\parallel} w_0}& k_y\Delta_p \\
-k_x\Delta_p & -k_y\Delta_p & -\frac{\Delta_pk^2_{\parallel}}{w_0}
\end{array}\right)\nonumber\\ & & 
\ee
where $\gamma=\mathrm{sign}(z_i-z_j)$,
$\ve{k}_{\parallel}=m\ve{u}'+n\ve{v}'+\ve{k}_{0\parallel}
=k_x\ve{\hat{x}}+k_y\ve{\hat{y}}$, and $w_0$ is the component along
$z$ of the wave vector $\ve{k}_0$, i.e.,
$w_0=(k_0^2-k_x^2-k_y^2)^{1/2}$.  $\Delta_p$ and $\Delta_s$ are the
Fresnel reflection coefficients for the substrate.  Sums involving
different susceptibility tensors (free-space or surface) will have a
different behavior, due to the different arguments of the exponential
terms ($z_i+z_j$ and $|z_i-z_j|$).  They will be computed separately.

For the surface term, the convergence of the sum is ensured by the
exponential term. As $m$ and $n$ increase, the magnitude of
$\ve{k}_{\parallel}$ increases and the nature of the plane wave
changes from propagating to evanescent. Because $z_i+z_j$ never
vanishes, and because the subunits are never exactly on the surface,
this exponential term is always present and ensures the rapid
convergence of the sum.

For the free-space part, the argument of the exponential term is
$|z_i-z_j|$ and the rapid convergence of the sums is not as
trivial. We use the method introduced to derive the Green function of
a 2D square grating.~\cite{mittra} We consider two cases. The first
case pertains to the interaction between elements from different
``layers'' of the lattice, and corresponds to the case $z_i\neq
z_j$. This case is similar to the surface problem and the convergence
of the sum is ensured by the exponential term.

In the second case $z_i=z_j$ and the exponential term disappears. We
cast the free-space part of the infinite sum in two different
forms. We note by $\ve{a}(\ve{r}_{\parallel},z_i-z_j)$ the sum in the
direct space ($\ve{F}$ terms in Eq.(\ref{somme})). We note by
$\ve{A}(\ve{k}_{\parallel},z_i-z_j)$ the sum in the reciprocal space
($\ve{G}$ terms in Eq.(\ref{dipi8})). When $z_i=z_j$ we write the sum
as
\be
\label{sum0}
\ve{a}(\ve{r}_{\parallel},0) & = & \ve{A}(\ve{k}_{\parallel},h)
+[\ve{a}(\ve{r}_{\parallel},0)-\ve{a}(\ve{r}_{\parallel},h)], \ee
where $h$ is an offset parameter.  The auxiliary sum
$\ve{A}(\ve{k}_{\parallel},h)$ can be computed efficiently owing to
the presence of an exponentially decreasing term.  The difference of
direct-space sums
$\ve{a}(\ve{r}_{\parallel},0)-\ve{a}(\ve{r}_{\parallel},h)$ goes as
$1/r_{\parallel}^2$ and can also be computed efficiently.  With
Eq.(\ref{sum0}) we can ensure a rapid convergence of the sums in a
discretization plane despite the absence of an exponentially
decreasing term in the original sum.

To improve further the convergence of the sums, we use Shanks'
accelerator.~\cite{shanks} Because we have two sums (over $m$ and $n$)
one solution would be to apply successively Shanks' accelerator to the
inner ($n$) and outer ($m$) sums (as suggested in
Ref.~\ref{singh}). The problem with this approach is that in our case,
the convergence of the inner sum (over $n$) can be very slow for high
values of $m$ (outer sum). A better solution consists in defining one
element $l$ of the Shanks's series as the sum over $m=-l,l$ for
$n=-l,...,l$ and $n=-l,l$ for $m=-l+1,...,l-1$.  This strategy gets
rid of the inner/outer sum problem and results in a faster convergence
and an easier implementation of the Shanks algorithm.

Note that there is another way of computing efficiently the free-space
term. As we did earlier, when we introduced a parameter $h$, it is
possible to split the infinite sum ($\ve{F}$) terms in
Eq.(\ref{somme}) in two parts; one in the direct space and one in the
reciprocal space, where these two sums converge quickly owing to a
damping function.~\cite{poppe,rmq} The convergence is the best when
$h=\sqrt{\pi/(u_xv_y-v_xu_y)}$. Poppe {\it et al.}  introduced this
method to study the optical response of an atomic monolayer; the
period of the structure was therefore very small compared to the
wavelength.

Once the periodic susceptibility tensors are known, we solve the
linear system of Eq. (\ref{dipi4}) to find the self-consistent field
at each site.  Once the field at all subunits is known, the scattered
field at any position $\ve{r}$, above, below or inside the periodic
structures is readily computed through Eq.(\ref{dipi4}) with the
exchange $\ve{r}\leftrightarrow\ve{r}_i$.  Notice that the new linear
system is no longer symmetric.  This is due to the fact that the
elements of the system depend on the incident plane wave via the
exponential term in Eq.(\ref{dipi4}).

\section{Example: Scattering by a periodic structure lying on a substrate}

To illustrate the method we consider the case of a dielectric
substrate (the relative permittivity is 2.25) on which lies a 2D
grating of parallelepipeds with the same permittivity. The structure
is illuminated in TM polarization from the substrate side by total
internal reflection at an angle of incidence $\theta=45^{\circ}$; then
$\ve{k}_{0\parallel}=(\frac{2\pi}{\lambda} \sin \theta
\sqrt{2.25},0)$. The wavelength in vacuum is $\lambda=632.8$ nm, and
the basis vectors of the lattice $\ve{u}=(a,0)$, $\ve{v}=(0,a)$. 

\begin{figure}[H]
\begin{center}
\includegraphics*[draft=false,width=80mm]{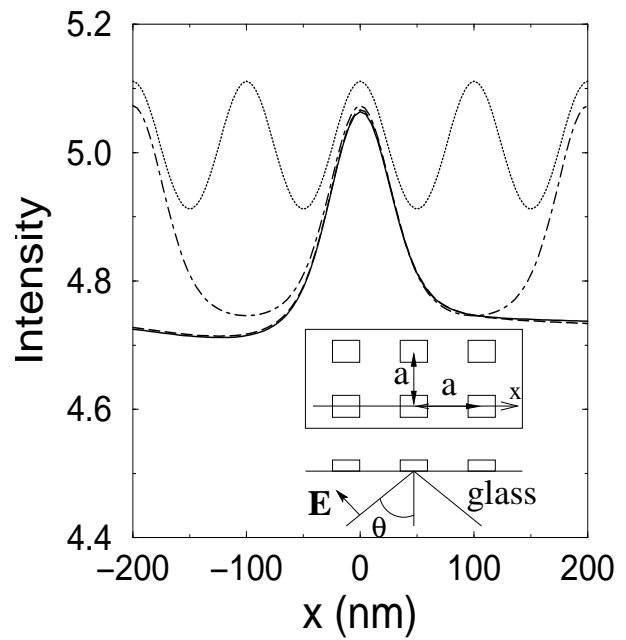}
\end{center}
\caption{Intensity of the electric field above a dielectric substrate
in the direction of the x-axis with a 2D grating of
parallelepipeds. The inset shows the geometry used. The solid line is
for an isolated parallelepiped.  The other curves are obtained for the
2D grating with $a=100$ nm (dotted line), $a=200$ nm (dot dashed
line), $a=1000$ nm (dashed line).}
\label{multig}
\end{figure}

The parallelepipeds have a square base of $40\times 40$ nm$^2$, and a
height of 20 nm (see inset Fig.~2). In Fig. 2 we present the intensity
of the electric field along the $x$-axis, 60 nm above the dielectric
substrate for different value of $a$.  The curves are obtained for
$N=256$, hence the size of the subunit is $5\times 5\times 5$ nm$^3$
(but convergence is already achieved for $N=32$). Notice that the
solid line is for an isolated parallelepiped on the substrate, i.e.,
the electric field is computed with the conventional
CDM.~\cite{chaumet1} When $a$ is small, the computed curves for the
electric field are notably different from the single object case. This
denotes a strong coupling between parallelepipeds.  Conversely for
large $a$ ($a=1000$ nm), the curve is very similar to the curve for an
isolated parallelepiped.

Table 1 presents the computation time for the coefficients of the
linear system (Eq. (\ref{somme})) used to solve Eq. (\ref{dipi4}), for
different values of $N$, and three values of $a$. The factor $h$ has
an important influence on the computation time, therefore we have
chosen the optimal value of $h$ for each case. As a reference we use
the conventional CDM to compute the field for a single
parallelepiped.~\cite{computer}

\end{multicols}

\begin{table}
\begin{center}
\begin{tabular}{|c|c||c|c|c|c|}
\multicolumn{2}{|c||}{$N$} & 32 & 256 & 500 & 1372 \\
\hline
\multicolumn{2}{|c||}{CDM} & 2 &  18 &39 & 137 \\
\hline
$a=$ & CDM$_1$ & 0.3 (2)  & 4 (17) & 10 (34)  & 43 (116) \\
\cline{2-6}
 100 nm & CDM$_2$ & 0.2 (0.4)  & 2.7 (5.5)  & 7 (13) & 29 (54) \\
\hline
$a=$ & CDM$_1$ & 0.7 (2)  & 12 (30)  &  30 (75)  & 119 (300) \\
\cline{2-6}
 200 nm & CDM$_2$ & 0.4 (1) & 7 (16) & 18 (40) & 72 (158)  \\
\hline
$a=$ & CDM$_1$ & 5.7 (16)  & 96 (281)  & 246 (684)  & 949 (4020)  \\
\cline{2-6}
1 $\mu$m  & CDM$_2$ & 5.6 (16)  & 96 (276)  & 233 (674)  & 900 (2460)  \\
\end{tabular} 
\label{tasphere}
\end{center}  
\caption{Computation time in seconds for the coefficients of the
linear system (Eq. (\ref{somme})) used to solve Eq. (\ref{dipi4}).
$N$ is the number of subunits. CDM is the time for the classical CDM
for one parallelepiped. CDM$_1$ is the time for the periodic CDM when
the free space contribution is computed with Eq.(\ref{dipi8}), and
CDM$_2$ is the time for the periodic CDM when the free space
contribution is computed with Ref.[\ref{poppe}]. The infinite sums of
the series are stopped when the relative error is less than $10^{-3}$
($10^{-6}$).}
\end{table}

\begin{multicols}{2}

Table 1 shows three computation times: CDM is the time for the
classical CDM for one parallelepiped. CDM$_1$ is the time for the
periodic CDM when the free space contribution is computed with
Eq.(\ref{dipi8}), and CDM$_2$ is the time for the periodic CDM when
the free space contribution is computed with Ref.[\ref{poppe}].
CDM$_2$ is faster than CDM$_1$ for small periods. For $a=1 \mu$m the
computation times are similar. For larger periods CDM$_2$ fails to
converge to the reference result because the method of
Ref.[\ref{poppe}] used to compute the free space term does not work
well for large $a$.  We note that the computation time increases with
$a$. This is due mainly to the surface term. The convergence of the
series depends on the term $\exp[iw_0(z_i+z_j)]$. In our case the
modulus of the vectors of the reciprocal basis are
$|\ve{u}'|=|\ve{v}'|=2\pi/a$.  Hence when $a$ decreases, the modulus
of the vector basis increases, $w_0$ becomes imaginary for smaller
values of $(m,n)$, and the exponential term produces a stronger
damping. Obviously, when $N$ increases, the computation time increases
due to the increased number of subunits involved.  But there is
another effect of the surface term. As the size of the subunit becomes
smaller ($N$ increases), there are more subunits close to the
substrate with a small value of $z_i+z_j$ and a slower exponential
decay. When we compare the classical CDM to the periodic CDM we see
that for $a$ smaller than 200 nm the computation time of the periodic
CDM is shorter. When the size of the period becomes larger than the
wavelength used, the convergence becomes slower.

\section{Conclusion}
In conclusion we have generalized the coupled dipole method (CDM) to
periodic structures. We have discussed explicitly the case of a
three-dimensional structure, periodic in two directions, placed on a
substrate. However, the principle of the approach described here
applies to a broad range of configurations with one, two or
three-dimensional structures.  The main advantage of this new
formulation is that it relies on the same straightforward,
direct-space discretization scheme that is used for a single localized
object. Therefore, all the knowledge acquired previously in CDM
modeling of finite systems can be transposed to the study of periodic
structures.~\cite{salomon} Optical anisotropy, for instance, can be
included by taking the appropriate permittivity tensor. Also, as shown
here, the symmetry of the periodic lattice can be arbitrary.  Here, we
have considered the case of plane wave illumination. In the case of
arbitrary illumination, each spectral component of the incident field
must be treated individually.  An interesting extension of the present
work would be to merge the periodic CDM and the conventional CDM into
a single approach to light scattering. This would be particularly
useful in dealing with localized defects in periodic structures or the
interaction between a near-field probe (microscope tip, fluorescing
particle,...) and a periodic system.  The periodic generalization of
the coupled dipole method can also be used to draw a better physical
picture of the local-field corrections that appear during the multiple
scattering of light by a discrete set of scatterers.~\cite{rahmani3}

P. C. Chaumet's email address is pchaumet@loe.u-3mrs.fr.

\end{multicols}


\begin{thebibliography}{0}          

\bibitem{purcell:73} E. M. Purcell and C. R. Pennypacker,
Astrophys. J. {\bf 186}, 705 (1973).

\bibitem{draine:88}\label{draine} B. T. Draine, Astrophys. J. {\bf
333}, 848 (1988); B. T. Draine and J. Goodman, Astrophys. J. {\bf
405}, 685 (1993); B. T. Draine and P. J. Flatau , J. Opt. Soc. Am. A
{\bf 11}, 1491 (1994) and references therein.

\bibitem{schmehl:97} R. Schmehl, B. M. Nebeker, and E. D. Hirleman,
J. Opt. Soc. Am. A {\bf 14}, 3026 (1997).

\bibitem{chaumet1} \label{chaumet1} P. C. Chaumet, and
M. Nieto-Vesperinas, Phys. Rev. B. {\bf 61}, 14119 (2000); {\bf 62},
11185 (2000); {\bf 64}, 035422 (2001).

\bibitem{rahmani1} A. Rahmani, P. C. Chaumet, and F. de Fornel,
Phys. Rev A {\bf 63}, 023819 (2001).

\bibitem{rahmani2} A. Rahmani and G. W. Bryant, Opt. Lett. {\bf 25},
433 (2000).

\bibitem{agarwal} \label{agarwal} G. S. Agarwal, Phys. Rev. A {\bf
11}, 230 (1975); {\bf 12}, 1475 (1975).   

\bibitem{chaumet2} \label{chaumet2} P. C. Chaumet and
M. Nieto-Vesperinas, Opt. Lett.  {\bf 25}, 1065 (2000).

\bibitem{mittra} R. E. Jorgenson, and R. Mittra, IEEE Trans. Antennas
Propagat. {\bf 38}, 633 (1990).

\bibitem{shanks} D. Shanks, J. Math. Phys. {\bf 34}, 1 (1955).

\bibitem{singh}\label{singh} S. Singh and R. Singh , IEEE
Trans. Microwave Theory Tech. {\bf 39}, 1226 (1991).

\bibitem{poppe}\label{poppe} G. P. M. Poppe, C. M. J. Wijers, and
A. van Silfhout, Phys. Rev. B {\bf 44}, 7917 (1991).

\bibitem{rmq} Note that there are typographical errors in
Ref.[\ref{poppe}].  In Eq.(A15) the argument of the exponential term
of the first sum should have a plus sign. In Eq.(A19) the argument of
the erfc function should have a minus sign.

\bibitem{computer} We use a 750MHz monoprocessor PC.

\bibitem{salomon} L. Salomon, F. D. Grillot, A. V. Zayats, and F. de
Fornel, Phys. Rev. Lett. {\bf 86}, 1110 (2001);
L. Mart{\'{\i}}n-Moreno, F. J. Garc{\'{\i}}a-Vidal, H. J. Lezec,
K. M. Pellerin, T. Thio, J. B. Pendry, and T. W. Ebbesen,
Phys. Rev. Lett. {\bf 86}, 1114 (2001).

\bibitem{rahmani3} A. Rahmani, and G. W. Bryant, Phys. Rev. A {\bf
65}, 033817 (2002); A. Rahmani, P. C. Chaumet, and G. W. Bryant,
Opt. Lett. {\bf 27}, 430 (2002).

\end{thebibliography}
\end{document}